\documentclass[11pt,twoside]{article} 
\usepackage{times,fancyhdr}
\usepackage{natbib}
\usepackage{graphicx}

\def\ie{{i.e.}}
\def\eg{{e.g.}}
\def\etal{et al.}

\def\r1q{\mbox{$r^{1/4}$}}

\date{}
\title{Tidal energy effects of dark matter halos on early-type galaxies.} 
\author{Valentinuzzi T., Caimmi R., D'Onofrio, M.} 

\begin{document} 

\maketitle 

\pagestyle{fancy}
\fancyhead{} 
\fancyhead[EC]{Valentinuzzi \etal} \fancyhead[EL,OR]{\thepage}
\fancyhead[OC]{Tidal energy of dark matter halos}
\fancyfoot{} 
\renewcommand\headrulewidth{0.5pt}
\addtolength{\headheight}{2pt} 

\abstract Tidal interactions between neighboring objects span across
the whole admissible range of lengths in nature: from, say, atoms to
clusters of galaxies i.e. from micro to macrocosms. According to
current cosmological theories, galaxies are embedded within massive
non-baryonic dark matter (DM) halos, which affects their formation
and evolution. It is therefore highly rewarding to understand the
role of tidal interaction between the dark and luminous matter in
galaxies. The current investigation is devoted to Early-Type
Galaxies (ETGs), looking in particular at the possibility of
establishing whether the tidal interaction of the DM halo with the
luminous baryonic component may be at the origin of the so-called
``tilt'' of the Fundamental Plane (FP).  The extension of the tensor
virial theorem to two-component matter distributions implies the
calculation of the self potential energy due to a selected
subsystem, and the tidal potential energy induced by the other one.
The additional assumption of homeoidally striated density profiles
allows analytical expressions of the results for some cases of
astrophysical interest. The current investigation raises from the
fact that the profile of the (self + tidal) potential energy of the
inner component shows maxima and minima, suggesting the possible
existence of preferential scales for the virialized structure, \ie\
a viable explanation of the so called "tilt" of the FP. It is found
that configurations related to the maxima do not suffice, by
themselves, to interpret the FP tilt, and some other relation has to
be looked for.

\section{Introduction}

According to current cosmological theories, about $85$\% of existing
mass in the universe is in the form of (non baryonic) dark matter
(hereafter quoted as DM), whose tidal energy effects on the embedded
(baryonic) matter could be large. Since the first evidence of DM
presence in galaxy clusters \citep[\eg,][]{acta6-110}, the existence
of massive, non baryonic halos is consistent with present-day CMB
surveys, large scale galaxy clusters studies, and necessary, \eg,
for a viable explanation of flat rotation curves well outside
visible disks of spiral galaxies \citep[for an exhaustive review on
DM refer to,][]{dmreview}. The current investigation aims to provide
further insight on the tidal action induced by massive halos on
hosted galaxies, taking into consideration special sequences of
two-component systems, intended to model early-type galaxies
(hereafter quoted as ETGs) and their hosting halos.

The idea of exploring two-component systems, a stellar spheroid
completely embedded in a DM halo, moves from the fact that the
virial potential energy (hereafter quoted as VPE) of the stellar
component, shows a non-monotonic trend as a function of the radius,
as opposed to one-component systems. This behavior is induced by the
DM halo tidal potential, and is more effective for shallower DM halo
density profiles. The occurrence of extremum points in potential
energy could be highly rewarding, as in mechanics they correspond to
stationary points and may be special configurations for the system.
These extremum points could be a key to the explanation of the so
called ``tilt'' of the Fundamental Plane (FP)
\citep[see,][]{apj399-462}.

The current investigation is based on two ETGs density profiles of
astrophysical interest, using the formalism of the two-component
virial theorem for an explicit expression of the VPE of the stellar
subsystem embedded in the DM halo. The models and related special sequences of two-component systems,
intended to represent ETGs, are defined in Section 2.   An analysis
of VPE extremum points, with the further restriction of energy
conservation, is performed and discussed in Section 3.   Comparisons
between model predictions and both data from observations and results
from computer simulations, are made in Section 4. Conclusions are drawn in Section 5.
\\
\\
The complete article is part of the book entitled {\bf Energy Research Developments: Tidal Energy, Energy Efficiency and Solar Energy}, published by NOVA PUBLISHERS, and is available through the following web site: {\it https://www.novapublishers.com/catalog/product\_info.php?products\_id=7980}.

\end{document}